# A Moving-Object Index for Efficient Query Processing with Peer-Wise Location Privacy

Dan Lin
Computer Science
Missouri University of
Science & Technology
lindan@mst.edu

Christian S. Jensen
Computer Science
Aarhus University
csj@cs.aau.dk

Rui Zhang
Computer Science &
Software Engineering
The University of Melbourne
rui@csse.unimelb.edu.au

Lu Xiao
Computer Science
Missouri University of
Science & Technology
lx787@mail.mst.edu

Jiaheng Lu
Key Laboratory of Data
Engineering DEKE
Renmin University of China
jiahenglu@ruc.edu.cn

## ABSTRACT

With the growing use of location-based services, location privacy attracts increasing attention from users, industry, and the research community. While considerable effort has been devoted to inventing techniques that prevent service providers from knowing a user's exact location, relatively little attention has been paid to enabling so-called peer-wise privacy—the protection of a user's location from unauthorized peer users. This paper identifies an important efficiency problem in existing peer-privacy approaches that simply apply a filtering step to identify users that are located in a query range, but that do not want to disclose their location to the querying peer. To solve this problem, we propose a novel, privacy-policy enabled index called the *PEB-tree* that seamlessly integrates location proximity and policy compatibility. We propose efficient algorithms that use the PEB-tree for processing privacy-aware range and $k$NN queries. Extensive experiments suggest that the PEB-tree enables efficient query processing.

## 1. INTRODUCTION

We are experiencing an increasing availability of location-based services such as AT&T's TeleNav GPS Navigator, Sprint's Family Locator, and Intel's Thing Finder. A key obstacle to the broad adoption of location-based services is the lack of location privacy protection [2, 20, 30].

Specifically, in a setting where a service provider serves multiple users, a user may have privacy concerns with respect to the service provider as well as the other service users. As an example of the first case, a user may worry that the service provider will disclose the user's locations (e.g., the user's daily route) to malicious parties. We use *provider-wise privacy* for privacy in relation to the service provider. As an example of the second case, an employee may not want work colleagues to know his/her location during lunch if he/she is outside the company building. This type of access restriction may also prevent stalking or other personal security threats [24, 34]. We use *peer-wise privacy* for privacy in relation to peer users.

Most research on location privacy thus far has been devoted to provider-wise privacy, and techniques such as spatial cloaking [10, 36], location distortion [18], and encryption [9] have been explored. In relation to peer-wise privacy, only a simple filtering approach has been employed.

The setting of the filtering approach is one where users specify their privacy preferences using location privacy policies that capture who is allowed to see the location of who and under what conditions. To answer a peer-wise privacy-aware query, the filtering approach first finds users who satisfy the query's location requirements in the same way as is done for privacy unaware location-based queries, i.e., using existing moving object indexing and querying techniques. Only then it filters out users by inspecting their location privacy policies.

For example, if a user issues a query for other nearby service users, the service provider not only needs to find nearby users; it also needs to check the privacy policies of the users found to ensure that they are willing to disclose their locations to the querying user. When potential query results are found solely according to spatial proximity, which is well supported by existing indexing and query processing techniques, very large and unnecessary intermediate results may occur because the policy checking may eliminate most of the results. Section 3 further elaborates on the problem.

This paper aims to provide an indexing technique and accompanying query processing algorithms that enable the efficient processing of peer-wise privacy aware queries that serve as the foundation for typical location-based services. Our proposed approach is orthogonal to existing approaches to supporting provider-wise privacy and can be integrated with these to achieve additional privacy.

In particular, we propose the so-called Policy-Embedded $B^x$-tree (PEB-tree), which organizes objects based on both spatial proximity and privacy policy compatibility. The main idea is to generate an indexing key value for each object that encodes location as well as policy information. This way, objects spatially near each other and with compatible privacy policies are assigned similar keys and are placed near each other in the index. The PEB-tree is based on the widely implemented $B^+$-tree, which promises easy integration into existing commercial database systems. Based on the PEB-tree, we provide algorithms for processing privacy-aware range and $k$ nearest neighbor ($k$NN) queries.

The results of extensive empirical studies with the proposals suggest that the PEB-tree based algorithms outperform existing techniques considerably in terms of I/O cost.

The rest of the paper is organized as follows. Section 2 reviews related work. Section 3 gives problem definitions, and Section 4 describes the existing approach used for comparison. Section 5 presents the proposed policy-embedded indexing techniques along with a cost analysis. Then Sections 6 and 7 cover cost modeling and empirical performance studies, respectively. Section 8 concludes and outlines future research directions.

## 2. RELATED WORK

As the PEB-tree integrates moving-object location and privacy, we first discuss research in moving-object database management and then location privacy. After that, we review works that share concepts that underlie our work.

### 2.1 Indexing and Querying Moving Objects

*Previous Indexing Approaches*

Moving object indexing must contend with frequent updates. Thus, focus is often on the efficient support for workloads that contain queries as well very frequent updates, which contrasts earlier works on spatial indexing where the data was assumed to be relatively static and focus was on query performance.

Most recent indexing proposals fall into one of three main categories: (i) R-tree-based indexes, such as the RUM-tree [35], the TPR-tree [27], and the TPR*-tree [31]; (ii) B$^+$-tree-based indexes, such as the B$^x$-tree [13] and the B$^{dual}$-tree [32]; and (iii) quadtree-based indexes, such as STRIPES [25]. A benchmark study [3] finds that the TPR-tree, the B$^x$-tree, and STRIPES perform best under different workloads. However, these indexes focus on spatial proximity and offer no provisions for supporting privacy.

Two recent indexing proposals [4, 17] take into account both location proximity and text similarity for finding the top-$k$ most relevant spatial web objects. In particular, these leverage the inverted file for text similarity retrieval and the R-tree for spatial proximity querying. The PEB-tree also considers two aspects of the data it indexes, but it tackles a very different problem, privacy-concerned location-based queries.

Following other research in moving object databases [13, 27, 31, 32], we represent the position of a moving object as a linear function from time to point locations in two-dimensional Euclidean space: $\vec{x}(t) = \vec{x} + \vec{v}(t - t_u)$, where $\vec{x}$ and $\vec{v}$ are the two-dimensional position and velocity of the object at time $t_u$, and $t_u$ is the time of the most recent update. An object is thus given by the triple $(\vec{x}, \vec{v}, t_u)$.

An object issues a location update to the server when the deviation between its actual location and the predicted location based on its moving function exceeds a given threshold. Objects are required to issue an update at least once within a maximum update time $\Delta t_{mu}$ in order to keep the server informed about their existence.

We proceed to describe the B$^x$-tree that serves as the base structure of the PEB-tree.

*The B$^x$-Tree*

The B$^x$-tree is an efficient and practical moving object index [3] that exploits the B$^+$-tree, which renders it amenable to implementation in real database systems. To exploit the B$^+$-tree, the B$^x$-tree transforms the linear functions that capture object movements into single-dimensional values by means of a space-filling curve (e.g., the Z-curve) that is proximity preserving in the sense that points close to one another in 2-dimensional space tend to be close to one another in the transformed 1-dimensional space.

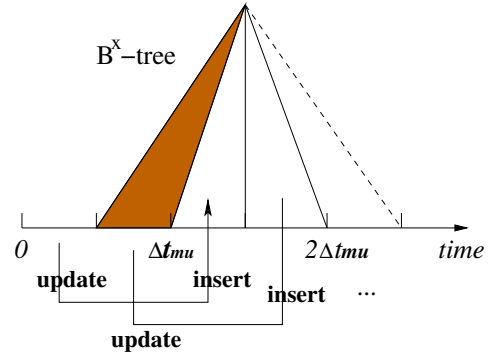

**Figure 1: Updates in the B$^x$-Tree**

To differentiate locations inserted at different times, the B$^x$-tree partitions the time axis into intervals of duration $\Delta t_{mu}/n$, where $\Delta t_{mu}$ is the maximum update interval and $n$ is a chosen number of sub-partitions within $\Delta t_{mu}$. Each partition has a label timestamp as shown in Figure 1. An update that occurs during some time interval is performed as of the nearest future label timestamp $t_{lab}$. This way, objects are assigned to different partitions of the time axis.

An object $O = (\vec{x}, \vec{v}, t_u)$, is then indexed as of $t_{lab} = \lceil t_u + t_{mu}/n \rceil_l$, where $\lceil x \rceil_l$ denotes the nearest later label timestamp of $x$. The value that is indexed, the $B^x value$, is the concatenation ($\oplus$) of the binary values ($[\cdot]_2$) of two components: the *index_partition*, computed from the label timestamp (Equation 2); and $x\_rep$, computed from the object location as of the label timestamp (Equation 3).

$$B^x value(O, t_u) = [index\_partition]_2 \oplus [x\_rep]_2 \quad (1)$$
$$index\_partition = (t_{lab}/(\Delta t_{mu}/n) - 1) \bmod (n+1) \quad (2)$$
$$x\_rep = x\_value(\vec{x} + \vec{v} \cdot (t_{lab} - t_u)) \quad (3)$$

For example, let the time axis be partitioned into intervals of duration $\Delta t_{mu}/2$. Objects updated between time 0 and $\Delta t_{mu}/2$ are indexed as of the time $t_{lab} = \Delta t_{mu}$. The resulting *index_partition* is 1 or '01' in binary format. Next, is the location as of $t_{lab}$ converted to a single-dimensional value using a space-filling curve. The B$^x$-tree inherits the B$^+$-tree's efficiency of insertions and deletions.

To process range queries using the B$^x$-tree, the query ranges need to be transformed to account for data transformation. Specifically, query ranges need to be enlarged to ensure that all objects that may be in the results are found.

Figure 2 shows an example where a solid rectangle $R$ is the query range at time 6 and black points are the locations of objects $A$, $B$, and $C$ as of time 5. Objects $A$ and $B$ will be in $R$ at time 6 according to their velocity vectors. To ensure that all objects are found, $R$ is expanded to $R'$ using the maximum object speeds along the two axes. For example, since the maximum downward speed is 2, the distance between the upper border of $R$ and $R'$ is obtained by multiplying this speed by the time difference, i.e., $2 \times (6-5) = 2$.

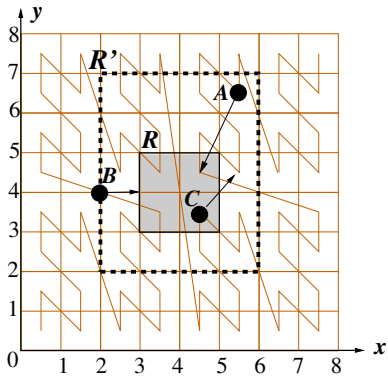

**Figure 2: Range Query in the $B^x$-Tree**

The enlargement guarantees that objects that may be in the result are found.

The enlarged query range is then converted into intervals of consecutive space-filling curve values. As a result, a sequence of range queries are issued to the $B^x$-tree. The objects found are then checked for inclusion in a refinement step by using their actual locations at the query time.

The $B^x$-tree can also process predictive $k$ nearest neighbor ($k$NN) queries. To do that, a range query based on an estimated $k$NN distance is issued first; then the range is enlarged gradually until $k$ nearest neighbors are found.

The PEB-tree augments the $B^x$-tree with privacy policy information and hence has novel policy encoding and index key generation algorithms. Moreover, the PEB-tree's query algorithm is more complex than that of the $B^x$-tree because both privacy and location proximity need to be considered simultaneously.

## 2.2 Location Anonymization

In provider-wise privacy, the service provider is typically prevented from knowing a user's exact location by using one or more of the following techniques: $k$-anonymization [29], spatial-temporal cloaking, and encryption. Gruteser et al. [10] are the first to apply $k$-anonymization to preserving location privacy and propose a spatial-temporal cloaking approach: For each user, a trusted third party (agent) generates a cloaking region in which at least $k-1$ other users are also present. The service provider receives regions instead of exact locations of users, and hence the service provider cannot distinguish a user from other users in the same region. Various extensions [1, 8, 15, 19, 21] aim to improve service flexibility and quality. A key limitation in these techniques is the performance bottleneck caused by the single anonymization agent. Further, the single agent can become a new target of attacks by malicious parties.

Next, several encryption-based location anonymization approaches [9, 14, 26] have been proposed. The most representative one is by Ghinita et al. [9], who employ Private Information Retrieval (PIR) to prevent service providers from knowing a user's location while providing a high quality of service.

Another thread of efforts [6, 12, 16] aims to perform location anonymization at the user side. However, this approach requires the users' devices to perform substantial computations and require extensive user involvement.

Despite extensive efforts on preserving provider-wise privacy, little work has appeared on peer-wise privacy protection. In Section 4, we cover two naive approaches to the indexing of moving objects with peer-wise privacy protection.

## 2.3 Additional Related Techniques

Works on spatial-keyword querying (e.g., [4, 17]) may seem similar to our work since they also build an index for two aspects: location proximity and text similarity. However, text similarity and privacy policy compatibility are very different. In addition, we consider moving objects, while spatial keyword querying indexes consider stationary.

We use a simple format for location privacy policy specification, which, however, contains the common major components of existing location privacy policy specifications [11, 23, 28]. Last, it is worth noting that location privacy policies are different from the concept of location-based access control, such as GEO-RBAC [5], in the sense that location data plays different roles. In location-based access control, location data serves as a condition that needs to be verified before a user is granted a permission to particular resources (e.g., classified documents), while location data is the data to be protected by location privacy policies.

## 3. PROBLEM DEFINITION

As mentioned, we represent the position of a moving user as a linear function from time to point locations in two-dimensional Euclidean space. The model enables the answering of queries on near future positions if needed, and the parameters needed for the use of this model are readily available from GPS receivers.

Next, we assume users will predefine their location privacy policies and that the server has access to all users' privacy policies. We define a succinct yet expressive format for Location-Privacy Policies (LPP for short) as follows.

DEFINITION 1. *Let $u_1$ and $u_2$ be two users. Let $P_{1\to 2}$ denote a location privacy policy assigned by $u_1$ for $u_2$. $P_{1\to 2}$ consists of three components $\langle role, loc_r, t_{int} \rangle$ given as follows.*

- *role: the relationship between $u_1$ and $u_2$, such as "family_member," "friend," or "colleague."*
- *$loc_r$: a spatial region.*
- *$t_{int}$: a subset of the time domain.*

A policy $P_{1\to 2} = \langle role, loc_r, t_{int} \rangle$ states that if $u_2$ is related to $u_1$ by relationship $role$ then $u_2$ is allowed to see $u_1$'s location when $u_1$ is located in $loc_r$ during $t_{int}$.

For example, Bob lets his colleagues see his location when he is in town (e.g., Chicago) during work hours (e.g., 8 a.m. to 5 p.m.). The corresponding LPP is: $P = \langle colleague, Chicago, [8\ a.m., 5\ p.m.] \rangle$. This way, access to Bob's location by users who are identified as colleagues by Bob is regulated by $P$. The use of the concept of $role$ is inspired by Role-based Access Control [7], which avoids writing the same policy for multiple people with the same relationship to Bob.

The specific design of the privacy policy format is orthogonal to the paper's contribution, which supports a range of spatio-temporal policy formats.

We support privacy-aware counterparts of the two arguably most fundamental query types, namely range queries and $k$ nearest neighbor queries. The formal definitions are given next.

DEFINITION 2. *(PRQ) The privacy-aware range query is defined as $PRQ = (qID, R, t_q)$, where $qID$ is the query issuer's identity, $R = ([x_1^l, x_1^u], [x_2^l, x_2^u])$ ('l' denotes 'lower bound' and 'u' denotes 'upper bound'), and $t_q$ is the query time. The query retrieves all users who satisfy the following two conditions: (1) the user's location $(x, y)$ at time $t_q$ falls within the query rectangle $R$; (2) the user has a location privacy policy $\langle role, loc_r, t_{int} \rangle$, in which $qID \in role$, $(x, y) \in loc_r$, and $t_q \in t_{int}$.*

In Definition 2, the condition $qID \in role$ checks if the relationship between the query issuer and the user is defined in the user's location privacy policy.

DEFINITION 3. *(PkNN) The privacy-aware k nearest neighbor query is defined as $PkNN = (qID, qLoc, k, t_q)$, where $qID$ is the query issuer's identity, $qLoc$ and $k$ nearest neighbor query parameters, and $t_q$ is the query time. The query retrieves $k$ users in $U$ for which no other users are nearer to the query issuer's location $qLoc$ at query time $t_q$, where $U$ is the set of all ($m > k$) users who have a location privacy policy $\langle role, loc_r, t_{int} \rangle$, where $qID \in role$, the user's location at time $t_q$ belongs to $loc_r$, and $t_q \in t_{int}$.*

To illustrate the problem that we tackle, we use the running example shown in Figure 3. The black point denotes a user with ID $u_1$ who wants to find her nearest friend. The star symbols represent $u_1$'s friends, whose IDs are $u_{12}$, $u_{30}$, $u_{59}$, $u_{100}$, and $u_{130}$. White circles represent other users. User $u_1$'s friends may have different location privacy policies. Suppose that at the time $u_1$ issues a privacy-aware nearest neighbor query, only one friend, i.e., $u_{12}$ (highlighted by the solid star symbol), is willing to disclose their location to $u_1$. The query result is then $\{u_{12}\}$.

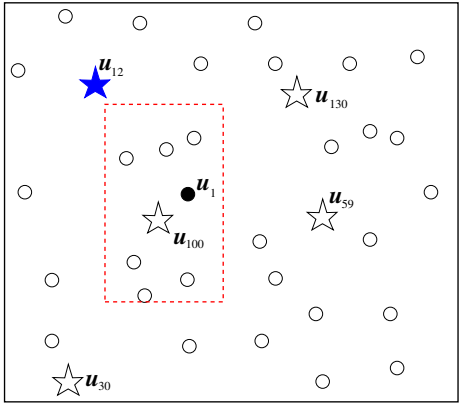

**Figure 3: Running Example**

## 4. SPATIAL INDEXING APPROACH

An existing approach [19] applies filtering to the result obtained from using a spatial index. In particular, the service provider processes the privacy-aware queries as were they normal spatial queries and then evaluates the privacy policies on the returned results. With this approach, many non-qualifying preliminary results may be retrieved from the spatial index.

A possible spatial index for the example scenario is given in Figure 4. Here, users are arranged purely based on their spatial proximity. For instance, $u_1$ and $u_{100}$ are stored together as they are close to one another.

To answer the privacy-aware nearest neighbor query from before, the service provider first locates $u_1$'s nearest neighbor $u_{100}$ and then evaluates $u_{100}$'s privacy policy with respect to $u_1$. Since $u_{100}$ does not allow $u_1$ to see his location at the query time, the service provider has to look for other nearby users. The query then needs to examine the next nearest neighbor, and this must be repeated until the final answer $u_{12}$ is found. In the example, at least four index nodes are accessed.

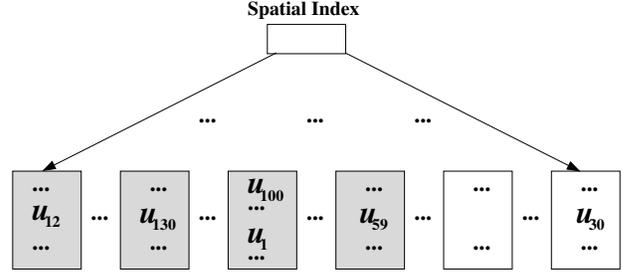

**Figure 4: Spatial Index Example**

## 5. POLICY-EMBEDDED B$^x$-TREE

To efficiently support privacy-aware queries, we propose a three-step approach. First, we develop a generic policy encoding technique that captures the compatibilities among location privacy policies belonging to different users. The encoded values are called *sequence values*. Second, we construct the Policy-Embedded B$^x$-tree (PEB-tree) that indexes mobile users according to both spatial and privacy policy proximity by carefully integrating sequence values with location mapping values. Third, we propose efficient algorithms for the queries defined in Section 3.

### 5.1 Location Privacy Policy Encoding

The actual policy encoding is preceded by policy translation and policy comparison phases.

In policy translation, the semantic locations defined in an LPP are mapped to Euclidean regions. In policy comparison, we use a score $\alpha \in [0, 1]$ to quantify the relationships between two users $u_1$ and $u_2$. If no location privacy policy is defined between $u_1$ and $u_2$, $\alpha = 0$; otherwise, $\alpha$ is determined by the size of the region and the duration of the time interval during which the two users allow each other to access their location information. If two policies are incompatible, $\alpha = 0$. As before, let $P_{1 \to 2}$ denote $u_1$'s policy regarding $u_2$. We consider two cases.

- $P_{1 \to 2} \leftrightarrow P_{2 \to 1}$: $u_1$ and $u_2$ are willing to simultaneously disclose their locations to each other under certain conditions. Thus, overlaps exist between the $loc_r$ and $t_{int}$ in the two policies. Let $O(loc_{r_1}, loc_{r_2})$ denote the area of the overlap between the two regions and let $D(t_{int_1}, t_{int_2})$ denote the duration of the overlap between the time intervals in the two policies. We define $\alpha$ for this case as follows, where the area $S$ of the space domain and the duration $T$ of the time domain are used for normalization.

$$\alpha = \frac{O(loc_{r_1}, loc_{r_2})}{S} \cdot \frac{D(t_{int_1}, t_{int_2})}{T}$$

- $P_{1 \to 2} \not\leftrightarrow P_{2 \to 1}$: $u_1$ and $u_2$ will not simultaneously disclose their locations to one another. In this case, at least one of $loc_r$ and $t_{int}$ in the two policies do not intersect. The corresponding $\alpha$, which never exceeds 0.5, is defined as follows.

$$\alpha = \frac{1}{2}\Big(\frac{|loc_{r_1}|}{S} \cdot \frac{|t_{int_1}|}{T} + \frac{|loc_{r_2}|}{S} \cdot \frac{|t_{int_2}|}{T}\Big)$$

The above function is also applied to the situation where only one user has a policy regarding the other. For example, if $P_{2 \to 1}$ does not exist, the second term in the definition is omitted.

It is worth noting that the above equations can be extended to cover the case where multiple policies exist between two users. Also, other policy comparison approaches may be adopted to compute $\alpha$ values.

Based on the obtained $\alpha$, we define the degree of compatibility between two users' policies, denoted as $C(u_1, u_2)$.

$$C(u_1, u_2) = \begin{cases} \frac{1}{2}(1+\alpha) & P_{1\to 2} \leftrightarrow P_{2\to 1} \\ \alpha & P_{1\to 2} \not\leftrightarrow P_{2\to 1} \\ 0 & \alpha = 0 \end{cases} \quad (4)$$

The compatibility function $C(\cdot, \cdot)$ returns a value in $[0, 1]$. The value is always greater than 0.5 for the first case, and it never exceeds 0.5 for the second case. The goal is to give higher priority to users who can sometimes see each other simultaneously than to users who always disclose their locations to one another under disjoint conditions. This is because two users belonging to the first case are more likely to be included in each other's query results. We call users with non-zero compatibility values *related users*. Otherwise, they are called *unrelated users*.

The next step is to determine the order of the sequence value assignment. We sort users in descending order of the number of their related users. This order gives higher priority to larger groups of users so as to preserve more relationships among users.

From the sorted list, we assign the first user, $u_1$, a sequence value $SV(u_1) = sv$ $(sv > 1)$. Each user $u_j$ related to $u_1$ obtains a sequence value $SV(u_j) = SV(u_1) + (1 - C(u_1, u_j))$. This scheme gives close sequence values to users with high compatibility values.

In what follows, only users who do not have a sequence value are considered. In particular, we select from the sorted list the next user $u_2$ and assign it a sequence value $SV(u_2) = SV(u_1) + \delta$, where $\delta > 1$. Parameter $\delta$ is an interval that helps separate different groups of users as well as leaves adjustment space for future policy updates. Then, each user $u_k$ related to $u_2$ obtains a value $SV(u_k) = SV(u_2) + (1 - C(u_2, u_k))$. This process continues until all users have sequence values. Policy updates are usually infrequent, and hence policy encoding is conducted largely off-line and does not add overhead at runtime.

Figure 5 outlines the algorithm of the sequence value assignment. First (lines 1–5), for each user $u_i$ in $U$, we put the related users (e.g., compatibility value $C$ is larger than 0) in the group $G(u_i)$. Then we sort the users in descending order of their group sizes and let $u_i$ be the $i$'th element of this list. After that, we start assigning sequence values for each user (lines 6–12). If a user $u_k$ has not been assigned a sequence value, the user obtains a sequence value that is $\delta$ larger than that of its predecessor. Next, we assign sequence values to all the group members of user $u_k$. Each group member without a sequence value obtains a sequence value equal to the sum of user $u_k$'s sequence value and the compatibility score with user $u_k$.

To illustrate the algorithm, we step through an example. Let 6 users $u_1, u_2, ..., u_6$ be given. Let their compatibility values be: $C(u_2, u_1) = 0.4$, $C(u_4, u_1) = 0.9$, $C(u_4, u_3) = 0.8$, $C(u_5, u_3) = 0.2$, $C(u_6, u_3) = 0.6$. According to the number of related users, we obtain this sorted list: $(u_3, u_1, u_4, u_2, u_5, u_6)$. Let the initial sequence value be 2 and also let $\delta = 2$. We first assign $u_3$ sequence value 2. Its related users $u_4$, $u_5$, and $u_6$ obtain the sequence values 2.2, 2.8, and 2.4, respectively. The next unassigned user is $u_1$ whose sequence value is set as follows: $CV(u_1) = SV(u_3) + \delta = 2 + 2 = 4$. User $u_2$ is currently unassigned and is related to $u_1$. Thus, $SV(u_2) = 4 + (1 - 0.4) = 4.6$. This completes the assignment.

## 5.2 Structure of the PEB-Tree

The PEB-tree is based on the $B^x$-tree [13], which in turn is based on the $B^+$-tree. This arrangement aims to make the PEB-tree easy to implement in real database management systems that invariably support $B^+$-trees.

A leaf node in the PEB-tree has the following format:

$$\langle PEB\_key, UID, x, y, v_x, v_y, t, Pnt_p \rangle,$$

where $PEB\_key$ is the index key, $UID$ is the user ID, $(x, y)$ and $(v_x, v_y)$ record the user's location and velocity at time $t$, and $Pnt_p$ links to the user's privacy policy set and other user-specific information. The internal nodes of the PEB-tree serve as a directory that contains index key values and pointers to child nodes.

The critical issue in building the PEB-tree is the generation of the $PEB\_key$ values for the users. A $PEB\_key$ consists of three components: (i) $TID$, which indicates the time partition in the PEB-tree in which a user's information is stored; (ii) $ZV$, which is the Z-curve [22] value of a user's location as of the time of the time partition $TID$; and (iii) $SV$, which is the policy encoding detailed in Section 5.1. The first two components are computed in a similar way as in the $B^x$-tree [13]. After we obtain the three components, we combine them as follows to form the $PEB\_key$.

$$PEB\_key = [TID]_2 \oplus [SV]_2 \oplus [ZV]_2 \quad (5)$$

Here, $[x]_2$ again denotes the binary value of $x$ and $\oplus$ denotes concatenation. The construction of the $PEB\_key$ gives higher priority to sequence values than to location mapping values. This design is attractive because users related to the query issuer are usually much fewer than the unrelated users within the vicinity of a query. Using the $PEB\_key$, users who have policies related to one another will tend to be stored close to each other, which reduces the cost of processing privacy-aware queries.

The algorithms for insertion and deletion of objects in the PEB-tree are similar to those for the $B^+$-tree. Each insertion or deletion requires only a single-path travel of the index, and the PEB-tree has similarly efficient update performance as the $B^+$-tree.

Figure 6 shows an expected PEB-tree that corresponds to the example from Section 4. The figure suggests that the PEB-tree arranges objects so that queries need fewer node accesses.

---

**Algorithm Sequence_Value_Assignment**
Output: assignment result $SV$
1.  for $i \leftarrow 1$ to $|U|$ do
    // $U$ is the list of all of users; $U[i] = u_i$
2.   $G(u_i) \leftarrow \emptyset$; $SV(u_i) \leftarrow \perp$
3.   for $j \leftarrow 1$ to $|U|$ do
4.    if $C(u_i, u_j) > 0$ then $G(u_i) \leftarrow G(u_i) \bigcup \{u_j\}$
5.  $U_l \leftarrow \text{Sort}(U, |G|, \text{desc})$
    // list $U_l$ contain users in descending order of $|G|$; $U_l[i] = u_i$
6.  $SV(u_1) \leftarrow sv$
7.  for $k \leftarrow 1$ to $n$
8.   if $SV(u_k) = \perp$ then
9.    $SV(u_k) \leftarrow SV(u_{k-1}) + \delta$
10.  for each $u_j$ in $G(u_k)$ do
11.   if $SV(u_j) = \perp$ then
12.    $SV(u_j) \leftarrow SV(u_k) + (1 - C(u_k, u_j))$
13. return $SV$

**Figure 5: Sequence Value Assignment**

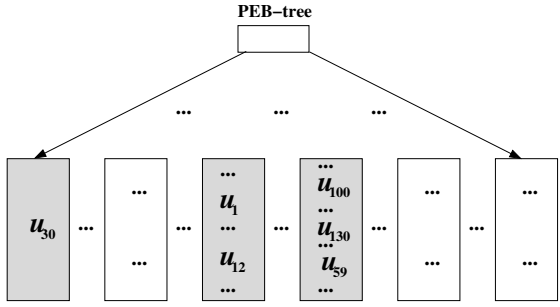

**Figure 6: PEB-Tree Example**

## 5.3 Privacy-Aware Range Query

The privacy-aware range query (PRQ, defined in Section 3) aims to find users who satisfy not only spatial constraints, but also policy constraints. To answer such a query, we first determine the search ranges for the two constraints separately and then combine them to form ranges that can be represented by $PEB\_key$ values. The query algorithm consists of four steps.

The first step finds all users in the query range. Let $U_{loc}$ denote the set of such users. The basic idea is similar to the range query in the $B^x$-tree [13]. Specifically, in each time partition $TID$, the query range $R$ is enlarged to cover users who are not in $R$ as of the time that they are indexed, but that may be in $R$ as of the query time. Then the enlarged query is converted into a set of one-dimensional intervals that are the search ranges of consecutive $ZV$ values. Let there be $k$ such intervals, given as follows: $\{[ZV_{s_1}; ZV_{e_1}], \ldots [ZV_{s_k}; ZV_{e_k}]\}$.

The second step finds the set of users (called $U_{pol}$) who may allow the query issuer to see their locations at the query time. For this purpose, we maintain a list for each user that stores the $SV$ values of users who have policies with respect to the list owner. Such lists are updated only rarely, e.g., when a user is blocked by a previous friend or when a user adds a new friend. We arrange the users with policies with respect to the list owner in an ascending order of their $SV$ values and denote the minimum and maximum $SV$ values by $SV_{min}$ and $SV_{max}$, respectively.

The third step computes the $PEB\_key$ range corresponding to the intersection of $U_{loc}$ and $U_{pol}$ as obtained from the previous steps. We first combine the starting and ending points of the $ZV$ ranges with each $SV$ value, which yields these search ranges:

$$[SV_{min} \oplus ZV_{s_1}; SV_{min} \oplus ZV_{e_1}],$$
$$[SV_{min} \oplus ZV_{s_2}; SV_{min} \oplus ZV_{e_2}],$$
$$\ldots \ldots,$$
$$[SV_{max} \oplus ZV_{s_k}; SV_{max} \oplus ZV_{e_k}].$$

Then we convert theses into intervals of consecutive $PEB\_key$ values by adding the $TID$ of the time partition under consideration.

The $PEB\_key$ ranges are used to retrieve the query results in the PEB-tree. During the search, once a candidate user is found, the remaining search intervals formed by this user's $SV$ value are skipped. Each candidate user's actual locations and policies are evaluated. If a user is verified to be the final result, all the remaining search intervals involving this user's $SV$ value are skipped.

Figure 7 summarizes the main steps of the range query algorithm. At the beginning, we find the minimum and maximum sequence values in the query issuer's friend list. We start considering the first time partition in the PEB-tree by setting $next\_timestamp$ to 0. For each time partition, we enlarge the original query range using the Enlarge() function. The obtained enlarged query window $R'$ is converted into a set of 1-dimensional intervals by ZVconvert() according to the Z-curve mapping. By concatenating the $TID$s (computed from $next\_timestamp$), the sequence values, and the $ZV$ values, we obtain the search range for the $PEB\_key$ values which is $[StartPnt; EndPnt]$. Then we locate the leaf node $current\_leaf$ that contains the starting point of the search interval, and we keep retrieving the right sibling nodes until the end of the search interval. The search stops after all $n$ time partitions are checked.

Since the calculation of $PEB\_key$ values uses interleaving algorithms, it is possible that the $PEB\_key$ intervals computed above overlap with one another. To avoid duplicate search, the $PEB\_key$ intervals are refined into a set of non-overlapping intervals that are then used for search in the PEB-tree.

We proceed to compute search ranges for the example in Figure 3. Assume that the dashed rectangle is the range querying for user $u_1$ to find his nearby friends, where the query range $R = ([2,2],[4,6])$. Suppose that the $SV$ values of $u_1$ and the friends are the following: $SV(u_1) = 46$, $SV(u_{12}) = 50$, $SV(u_{30}) = 25$, $SV(u_{59}) = 89$, $SV(u_{100}) = 55$, $SV(u_{130}) = 80$. For simplicity, we assume that the space is 8×8. Then $R$ is converted into two one-dimensional intervals according to the Z-curve mapping: [13; 16] and [25; 28]. Combining $SV$ and $ZV$, we obtain 10 search ranges for each $TID$. The following are the ranges for $TID = 0$:

- $[TID \oplus SV(u_{30}) \oplus ZV_{s_1}; TID \oplus SV(u_{30}) \oplus ZV_{e_1}]$
  $= [0 \oplus 25 \oplus 13; 0 \oplus 25 \oplus 16] = [1613, 1616]$

- $[TID \oplus SV(u_{30}) \oplus ZV_{s_2}; TID \oplus SV(u_{30}) \oplus ZV_{e_2}]$
  $= [0 \oplus 25 \oplus 25; 0 \oplus 25 \oplus 28] = [1625, 1628]$

- $[TID \oplus SV(u_{12}) \oplus ZV_{s_1}; TID \oplus SV(u_{12}) \oplus ZV_{e_1}]$
  $= [0 \oplus 50 \oplus 13; 0 \oplus 50 \oplus 16] = [3213, 3216]$

---

**Algorithm PRQ ($q$, $t_q$, $uid$, $friend\_list$)**
Input: $R$ is the query range and $t_q$ is the query time
      $uid$ is the ID of the user who issues the query
      $friend\_list$ is the list of users related to $uid$
Output: $result\_list$

1. $SV_{min} \leftarrow$ smallest sequence value in $friend\_list$
2. $SV_{max} \leftarrow$ largest sequence value in $friend\_list$
3. $next\_timestamp \leftarrow 0$
4. $more \leftarrow true$
5. while $more$
6.     $R' \leftarrow$ Enlarge($next\_timestamp$, $R$, $t_q$)
7.     $ZV\_intervals \leftarrow$ ZVconvert($R'$)
8.     for each $(ZV_{start}; ZV_{end})$ in $ZV\_intervals$
9.         $StartPnt \leftarrow TID \oplus SV_{min} \oplus ZV_{start}$
10.        $EndPnt \leftarrow TID \oplus SV_{max} \oplus ZV_{end}$
11.        $current\_leaf \leftarrow$ leaf node containing $StartPnt$
12.        for each user $u$ in $current\_leaf$ do
13.           if $u$ passes location and policy evaluation then
14.             add $u$ to $result\_list$
15.        if $current\_leaf$ contains $EndPnt$ then
16.           $next\_timestamp \leftarrow next\_timestamp + 1$
17.        else
18.           $current\_leaf \leftarrow current\_leaf$.right_sibling
19.     if $next\_timestamp \geq n \vee current\_leaf = \bot$ then
20.        $more \leftarrow false$
21. end while
22. return $result\_list$

**Figure 7: Algorithm for the Privacy-Aware Range Query**

- $[TID \oplus SV(u_{12}) \oplus ZV_{s_1}; TID \oplus SV(u_{12}) \oplus ZV_{e_1}]$
  $= [0 \oplus 50 \oplus 25; 0 \oplus 50 \oplus 28] = [3225, 3228]$

  ... ...

- $[TID \oplus SV(u_{59}) \oplus ZV_{s_1}; TID \oplus SV(u_{59}) \oplus ZV_{e_1}]$
  $= [0 \oplus 89 \oplus 25; 0 \oplus 89 \oplus 28] = [5725, 5728]$

During the search of these ranges, once a user is found in the first spatial range [13;16], the second range will be skipped since a user has only one location.

## 5.4 Privacy-Aware $k$NN Query

The algorithm for the privacy-aware $k$NN (P$k$NN, defined in Section 3) query is derived from the B$^x$-tree's privacy-unaware $k$NN query algorithm [13], which is answered by iteratively performing range queries with an incrementally enlarged search region until $k$ answers are obtained. First, a range $R_{q1}$ centered at $q$ and with radius $r_q = D_k/k$ is constructed, where $D_k$ is the estimated distance between the query issuer and its $k$'th nearest neighbor; $D_k$ can be estimated by the following equation, where $N$ is the total number of users [33]:

$$D_k = \frac{2}{\sqrt{\pi}} \left[ 1 - \sqrt{1 - \left(\frac{k}{N}\right)^{\frac{1}{2}}} \right]$$

Since a user location that is inserted at a certain time is stored in the index as of a future label timestamp, $R_{q1}$ is enlarged to $R'_{q1}$ similarly to what we did for range queries to cover all users who may be in $R_{q1}$ as of the query time. If at least $k$ users are currently covered by the inscribed circle of $R'_{q1}$ at time $t_q$, the $k$NN algorithm returns $k$ users and stops.

Otherwise, $R_{q1}$ is extended by $r_q$ to obtain $R_{q2}$ and the corresponding enlarged window $R'_{q2}$. This time, the region $R'_{q2} - R'_{q1}$ is searched. The process is repeated until $k$ users are found within the inscribed circle of the current range. During the search, the corresponding two-dimensional ranges are converted into a set of intervals in the transformed, one-dimensional space.

To answer the P$k$NN query, we need to consider the search ranges of both the $ZV$ and the $SV$ values for each time partition $TID$. The $ZV$ ranges determine the locations of the users who are close to the query issuer, which can be obtained by the general approach already covered, but with the following modification. For each query range, we consider only the one interval formed by the minimum and maximum 1-dimensional values of the query range.

The reason for this difference is the following. The P$k$NN query requires multiple rounds of range queries, and the corresponding 1-dimensional query intervals obtained from different rounds of enlargement may intersect. When we actually search those intervals in the index, it is possible that multiple query intervals are located in the same leaf node.

To avoid complex interval calculations and repeated leaf node accesses, we use a single query interval for each range query. Suppose that $n$ rounds of enlargement occur. For round $i$ (i.e., $R'_{qi}$), we denote the starting and ending points of the set of corresponding one-dimensional search intervals by $ZV_{s_i}$ and $ZV_{e_i}$, respectively. The ranges of $n$ rounds are given by: $\{[ZV_{s_1}; ZV_{e_1}], \ldots, [ZV_{s_n}; ZV_{e_n}]\}$.

The $SV$ ranges retrieve users who may be willing to disclose their locations to the query issuer. A smaller $SV$ value indicates that the corresponding user is more likely to disclose their location to the query issuer. Suppose that $m$ users are willing to let the query issuer see their locations under some conditions. By arranging these $m$ users in increasing order of their sequence values, we have the following list: $[SV(u_1), SV(u_2), ..., SV(u_m)]$.

$$\begin{bmatrix} [SV(u_1) \oplus [ZV_{s1};ZV_{e1}] & SV(u_1) \oplus [ZV_{s2};ZV_{e2}] \cdots SV(u_1) \oplus [ZV_{sn};ZV_{en}] \\ [SV(u_2) \oplus [ZV_{s1};ZV_{e1}] & SV(u_2) \oplus [ZV_{s2};ZV_{e2}] \cdots SV(u_2) \oplus [ZV_{sn};ZV_{en}] \\ \cdots & \cdots & \cdots & \cdots & \cdots & \cdots \\ [SV(u_m) \oplus [ZV_{s1};ZV_{e1}] & SV(u_m) \oplus [ZV_{s2};ZV_{e2}] \cdots SV(u_m) \oplus [ZV_{sn};ZV_{en}] \end{bmatrix}_{m \times n}$$

**Figure 8: Search Matrix**

where $SV(u_i)$ is the sequence value of user $u_i$.

Figure 8 shows the complete search space (represented as a matrix) in one time partition for a given P$k$NN query. The actual search is based on the values of the $PEB\_key$ computed from the $ZV$ and $SV$ ranges in each element of the matrix together with the $TID$ of the corresponding time partition.

The next step is to find a good search order to obtain the query result as soon as possible. Observe that ranges close to the upper-left corner of the matrix have shorter spatial distances to or closer $SV$ value differences from the query issuer. Therefore, those ranges are more likely to contain the final query results. Therefore, we apply a *triangular search order* as illustrated in Figure 9, where the arrows and numbers in the brackets define the search order. Following this order, the $ZV$ and $SV$ values are changed alternatively until $k$ candidates are found.

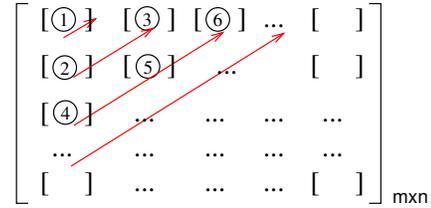

**Figure 9: Triangular Search Order**

Having found $k$ candidates, we check the remaining ranges in the last visited column in the search matrix, i.e., a vertical scan is done for the last visited column. For this vertical scan, the intervals of the $ZV$ values in the remaining ranges are shortened according to the distance to the latest $k$'th nearest candidate.

For example, if $k$ candidates are found after examining the users falling in the range indicated by the circle 5 (in the second column in Figure 9), we continue to consider the remaining ranges in the second column, which are: $[SV(u_2) \oplus [ZV_{s2}; ZV_{e2}], \ldots, SV_m \oplus [ZV_{s2}; ZV_{e2}]$. The interval of the $ZV$ values is a subset of $[ZV_{s2}; ZV_{e2}]$.

In particular, the new interval corresponds to the query square with the query issuer at the center and twice the distance to the $k$'th nearest candidate as its side length. This last step is needed in order to determine whether there are other users who have not been examined, but are closer to the query issuer than the $k$'th nearest neighbor found so far.

Figure 10 outlines the algorithm of the P$k$NN queries. First, we compute the estimated distance between the query issuer and the $k$'th nearest neighbor, based on which we obtain the initial query radius. The search starts from the first time partition in the PEB-tree, i.e., $next\_timestamp = 0$. In each time partition, the Get-Range() function constructs the search range which is a square centered at $(x, y)$ with length equal to $2r_q$. According to the search order adopted, the Next_friend() and Next_radius() functions compute the corresponding $SV$ value and radius of the query range, respectively. Theses parameters are then supplied to the PRQ query mod-

**Algorithm PKNN**($x, y, t_q, k, uid, friend\_list$)
Input: $(x, y)$ is user $uid$'s location
    $k$ is the required number of neighbors
    $t_q$ is the query time
    $friend\_list$ is the list of users related to $uid$
Output: $result\_list$

1. $D_k \leftarrow$ 2/sqrt(3.14) × (1- sqrt(1- sqrt(k/N)))
2. $r_q \leftarrow D_k/k$
3. $next\_timestamp \leftarrow 0$
4. $more \leftarrow true$
5. while $more$
6.     $R \leftarrow$ GetRange($(x, y), r_q$)
7.     $fid \leftarrow$ Next_friend($friend\_list$)
8.     $neighbor \leftarrow$PRQ($R, t_q, uid, fid$)
9.     $result\_list \leftarrow$Add_to_result($neighbor$)
10.    if $k$ neighbors are found
11.       $fid \leftarrow$ Rest_friend($friend\_list$)
12.       $R \leftarrow$ GetRange($(x, y), kdist$)
13.       $neighbor \leftarrow$ PRQ($R, t_q, uid, fid$)
14.       $result\_list \leftarrow$Add_to_result($neighbor$)
15.       $more \leftarrow false$
16.     $r_q \leftarrow$ Next_radius()
17. return $result\_list$

**Figure 10: Algorithm for the P$k$NN Query**

ule (presented in the previous section) to retrieve candidate query results.

The Add_to_result() function will verify the actual locations and policy constraints of the obtained results. If $k$ neighbors are found, the query range will be refined based on the distance between the query issuer and the $k$'th nearest neighbor found so far, and the range of the sequence value is refined by the Rest_friend() function that returns the list of $SV$ values in the last visited column in the search matrix. After refinement, another PRQ query is invoked to obtain the final query result. In case less than $k$ neighbors are found, the query radius is enlarged to start a new round of search.

## 6. QUERY I/O COST MODELING

In this section, we model the I/O cost of querying with the PEB-tree. We consider the privacy-aware range query as it is the most fundamental query.

The cost function we develop is based on the following assumptions on the datasets. To simulate different relationships among users, we first randomly divide users into groups and then generate policies for each user based on a parameter called the *grouping factor* ($\theta$) and defined as $\theta = \frac{N_{gr}}{N_p}$, where $N_{gr}$ is the number of policies that a user has regarding other users in the same group, and where $N_p$ is the user's total number of policies. The grouping factor ranges from 0 to 1. When the factor is 1, each user only has policies with users in the same group, and no policies connect users in different groups. When the factor is 0, there is no group, and each user may have policies with respect to any user in the system.

Our approach is to identify important parameters that significantly affect query performance and then integrate their effects into a cost function. Recall that the index keys in the PEB-tree are generated by incorporating the effects of policy compatibility and location proximity. Moreover, the policy compatibility is represented as a sequence value to which the location encoding is appended. As a result, the sequence value becomes the dominant factor during querying, while the location encoding provides only supplementary pruning. Thus, the cost function focuses on modeling the effect of the sequence value assignment on the query performance. An empirical validation (in Section 7.10) offers evidence that the approach yields a quite accurate cost function.

The sequence value assignment is determined by the grouping factor $\theta$, the number of policies per user (denoted as $N_p$), and the total number of users (denoted as $N$). When $\theta = 1$, the PEB-tree achieves the best performance. This is because when users are well grouped, query results are constrained to users that are stored together. The query cost increases when $\theta$ decreases. The worst-case scenario occurs when each user is allowed to have a policy with any other user in the system, i.e., $\theta = 0$. In this case, the sequence values fail to group users, as there are no groups. The I/O cost of a query is upper-bounded by the number of users related to the query issuer when each of the related users is stored in a different leaf node.

The above effect is modeled by the cost function $C_1$ in Equation 6, where $N_l$ is the total number of leaf nodes in the index; $N_p$ is the query cost in the above-mentioned worst-case scenario; and $N_p^\theta$ estimates the benefit obtained from grouping and captured by the grouping factor. The term 1 captures the minimum query cost when the query result is stored in one leaf node.

$$C_1 = \begin{cases} 1 + N_p - N_p^\theta & N_p \leq N_l \\ 1 + N_l - N_p^\theta & N_p > N_l \end{cases} \quad (6)$$

In summary, $C_1$ estimates the number of nodes needed for storing users related to the query issuer by taking into account $N_p$ and $\theta$.

Next, we consider the effect of the parameter $N$. A larger $N$ leads to larger groups of users being connected through policies. Since the sequence value assignment is conducted group-by-group in descending order of the group size, the existence of many larger groups tends to increase the distance among the sequence values belonging to two related users. In other words, it increases the probability that users in the same query result are stored in different nodes, which in turn increases the query cost.

The empirical studies covered in the next section show that the query cost increases linearly with $N$. Therefore, we model it as a linear function and integrate it into $C_1$ as follows.

$$C = \begin{cases} 1 + (a_1 \frac{N}{L^2} + a_2)(N_p - N_p^\theta) & N_p \leq N_l \\ 1 + (a_1 \frac{N}{L^2} + a_2)(N_l - N_p^\theta) & N_p > N_l \end{cases} \quad (7)$$

In Equation 7, $L$ is the side length of the space and $\frac{N}{L^2}$ is then the density of the object space. Parameters $a_1$ and $a_2$ are obtained by taking as input any two sample points (i.e., the query cost $C$) from the experiments on the datasets with the same location distribution. For example, $a_1 = 10$ and $a_2 = 0.3$, for data sets with uniform location distribution.

Using the cost function, we are interested in understanding the extents of the ranges of settings within which the PEB-tree is competitive. Specifically, we find that the PEB-tree performs worse than the spatial index approach described in Section 4 when each user is related to more than about 5% of all users. Considering a data set with 100K users, 5% is 5,000, which is already a large number of friends for a person.

Such a worst-case scenario may not occur in reality, as little privacy is actually achieved in such scenario. If all users are related to each other, every user grants some access to everyone else in the system. We believe that the general settings used in the empirical studies covered next, in which users tend to show certain privacy preference to a group of users, make more sense.

# 7. EMPIRICAL PERFORMANCE STUDY

Following a description of the settings of the study, we cover the offline cost of the initial index building. Then follows query performance studies where a range of workload settings are varied. We end with a cost model validation study.

## 7.1 Experimental Settings

We compare the performance of the PEB-tree with the approach of using spatial index as introduced in Section 4. Specifically, we select the $B^x$-tree [13] as the spatial index, and we adopt the commonly used filtering approach to handle peer-wise privacy concerns. Since the PEB-tree is based on the $B^x$-tree and the spatial indexing approach is based on the $B^x$-tree, the same settings from the literature [13], such as the number of tree partitions and the maximum update interval, are used for the two approaches.

We use two types of synthetic data sets of user positions, namely uniformly distributed positions and positions distributed in a spatial network, both in a space domain with area $1000 \times 1000$. In the uniform datasets, user positions are chosen randomly, and they move in randomly chosen directions and at speeds ranging from 0 to 3. One may think of the unit of space as being kilometers and the unit of speed as being kilometers per minute.

The network-based data sets are generated using an existing data generator [27], where users move in a network of two-way routes that connect a varying number of destinations. Objects start at random positions on routes and are assigned at random to one of three groups of objects with maximum speeds of 0.75, 1.5, and 3. Whenever an object reaches one of the destinations, it chooses the next target destination at random. Objects accelerate as they leave a destination, and they decelerate as they approach a destination.

In all datasets, for each user, we generate a given number of random policies by varying the spatial ranges and time intervals with respect to a set of other users. The relationships among users are modeled using the grouping factor introduced in Section 6. Unless stated otherwise, the dataset contains 60,000 uniformly distributed users, and each has 50 policies with a grouping factor of 0.7.

The default query window is quadratic with side length 200, and $k$ is 5 in the P$k$NN query. The parameters used are summarized in Table 1, where values in bold denote default values used.

The performance is evaluated in terms of I/O cost. The disk page size is set at 4K bytes, and a 50-page LRU buffer is simulated. We report only query performance as the two approaches achieve similarly good update performance.

**Table 1: Parameters and Their Settings**

| Parameter | Setting |
|---|---|
| Buffer | 50 pages |
| Number of users | 10K, 20K, ..., **60K**, ..., 100K |
| Maximum speed | 1, 2, **3**, 4, 5, 6 |
| Query window size | 100, **200**, ..., 1000 |
| $k$ ($k$NN query) | 1, ..., **5**, ..., 10 |
| Grouping factor ($\theta$) | 0 (uniform), ..., **0.7**, ..., 1.0 |
| Number of policies per user | 10, ..., **50**, ..., 100 |
| Number of destinations | **uniform**, 25, 50, 100, ..., 500 |

## 7.2 Preprocessing Time for Policy Encoding

In the first round of experiments, we study the preprocessing time used for policy encoding. This one-time processing is done offline when users are first registered.

Figure 11(a) shows the results when varying the total number of users from 10K to 100K. The experiments were conducted on a PC with a 2.53GHz Intel Xeon CPU and 4 Gbytes of memory. The time increases linearly with the number of users. We also observe that the preprocessing is very efficient, as it takes only about 10 seconds to compare location privacy policies and generate sequence values for 100K users.

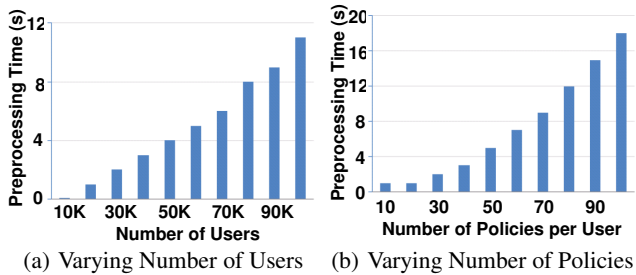

(a) Varying Number of Users    (b) Varying Number of Policies

**Figure 11: Preprocessing Time**

We also consider the policy encoding time when varying the number of policies to be analyzed for each user from 10 to 100, with 60K users. As shown in Figure 11(b), the processing time increases with the number of policies, but is still low. The efficient preprocessing can be attributed to our algorithm that uses the addition operation to directly generate sequence values related to a user instead of sorting compatibility degrees multiple times.

## 7.3 Effect of Total Number of Users

We proceed to evaluate the query performance of the PEB-tree and the spatial index approach. In this experiment, we vary the total number of users from 10K to 100K, and we measure the average I/O cost of 200 queries.

Figure 12(a) reports on privacy-aware range queries. We observe that the PEB-tree yields much less I/O than the spatial index. The performance gap increases with the data size. When the data size grows to 100K, the PEB-tree is about 10 times better than the spatial index. This behavior can be explained as follows. The spatial index organizes users only based on their spatial proximity. Thus, the spatial index needs to retrieve all users inside the query range, regardless of whether or not they are allowed to be seen by the query issuer, which increases costs. The PEB-tree stores users based on both location and policy proximity, and search is narrowed by using both location and policy constraints; hence it achieves the better performance.

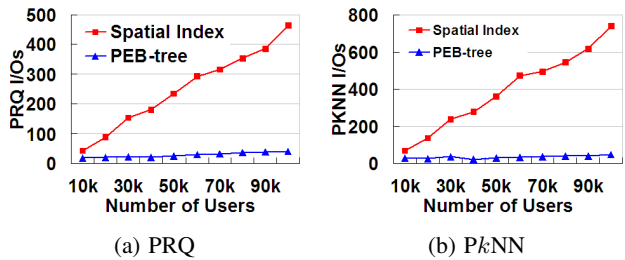

(a) PRQ    (b) P$k$NN

**Figure 12: Effect of Total Number of Users**

Figure 12(b) shows the performance of P$k$NN queries. Again, the PEB-tree significantly outperforms the spatial index approach. As for range queries, this demonstrates that the PEB-tree provides a better storage arrangement by considering both location and policy proximity, which in turn reduces unnecessary accesses to non-qualifying users.

The triangular search order, which examines users in descending order of their probabilities to be included in the query result, also improves performance. In other words, users who are either close to the query issuer or are more likely to be visible to the query issuer are checked early, which directs the search towards users who qualify for the result and shortens the query processing.

### 7.4 Effect of Number of Policies Per User

In this experiment, we vary the number of policies per user from 10 to 100. Without loss of generality, we assume that each user has only one location privacy policy with respect to a particular user.

Figure 13(a) shows the performance of privacy-aware range queries, from which we can see that the PEB-tree again outperforms the spatial index. Moreover, it is not surprising to observe an increase of the query cost in the PEB-tree with the number of policies. The more policies, the more qualified users may be included in a query result, and therefore more nodes are accessed. We also observe that the performance of the spatial index is independent of the number of policies. This is because the spatial index considers only location proximity. Thus, queries with the same location constraint will cause the same number of candidate users to be retrieved.

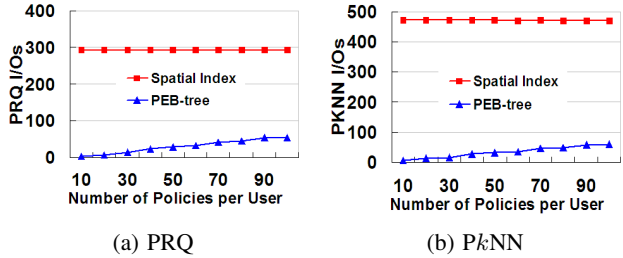

(a) PRQ  (b) P$k$NN

**Figure 13: Effect of Number of Policies per User**

Figure 13(b) compares the P$k$NN query performance of the two approaches. Observe that the PEB-tree saves significant I/O compared to the spatial index. The reason is similar to that discussed for the previous experiments.

### 7.5 Effect of Grouping Factor

Here, we investigate the effect of the grouping factor. As mentioned earlier, when this factor is 0, each user can have policies with randomly selected users in the system. When it is 1, each user is only related to users in the same group.

We first evaluate the range query performance. As shown in Figure 14(a), we can see that the cost of the PEB-tree tends to decrease as the grouping factor increases, whereas the spatial index maintains a constant performance. The experiment confirms the expectation that larger grouping factors help the PEB-tree achieve more effective sequence value assignments that group related users better. As the grouping factor approaches 1, users tend to be divided into non-overlapping groups. In this case, users in the same group are likely stored in the same or in a few nearby leaf nodes in the PEB-tree, and therefore few I/Os are needed for queries.

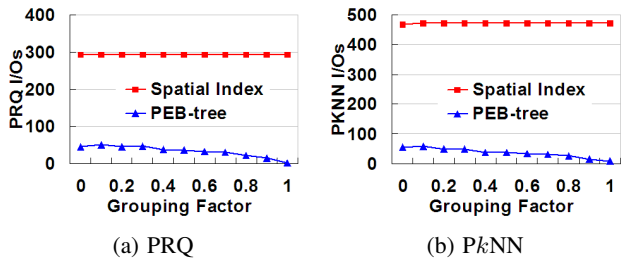

(a) PRQ  (b) P$k$NN

**Figure 14: Effect of the Grouping Factor**

However, the grouping factor does not influence the query performance of the spatial index since it stores users purely based on their location proximity, which is not influenced by the grouping factor.

Similar performance patterns are observed for P$k$NN queries, as shown in Figure 14(b). The PEB-tree performs the best for the same reasons.

### 7.6 Effect of Query Parameters

We now evaluate the impact of the location-related query parameters. For range queries, we measure the query cost by varying the query window side length from 100 to 1,000. For $k$NN queries, we vary parameter $k$ from 1 to 10.

Figure 15(a) shows the PRQ performance. Again, the PEB-tree significantly and consistently outperforms the spatial index. Moreover, the PEB-tree cost is almost constant, while the spatial index cost increases as the query window increases. The PEB-tree achieves constant performance because no matter how large the query window is, the maximum number of users to be checked by the PEB-tree is bounded by the total number of users related to the query issuer.

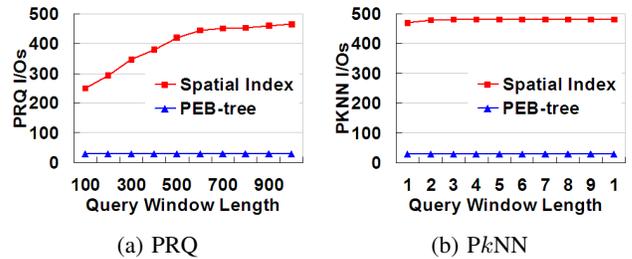

(a) PRQ  (b) P$k$NN

**Figure 15: Varying Query Parameters**

For the spatial index, the location-related query parameters play an important role. In particular, the larger the query window, the more nodes need to be accessed in the spatial index.

Figure 15 (b) shows the P$k$NN query performance of the two trees when varying $k$. The PEB-tree has stable performance for different values of $k$ due to the reasons similar to those stated for the last experiment. This also indicates that the PEB-tree is relatively unaffected by the location-related parameters. In the case of the spatial index, increasing the value of $k$ slightly degrades query performance since a larger $k$ requires the spatial index to enlarge the search range to find the qualified objects.

### 7.7 Effect of Spatial Distribution

This round of experiments targets the effect of the location distribution of the users. We observe the performance of range and nearest neighbor queries when using network-based data sets with the number of possible destinations (also called hubs) ranging from 25 to 500. The fewer the destinations, the more spatially skewed the data is.

Figure 16 shows that the PEB-tree achieves much better performance than the spatial index in all cases. The increase in the number of destinations only slightly affects the search ranges in the PEB-tree. This is because the location constraints are not the dominant factor during the index construction and hence has less influence on the query performance. The performance of the spa-tial index approach fluctuates slightly when varying the number of possible destinations.

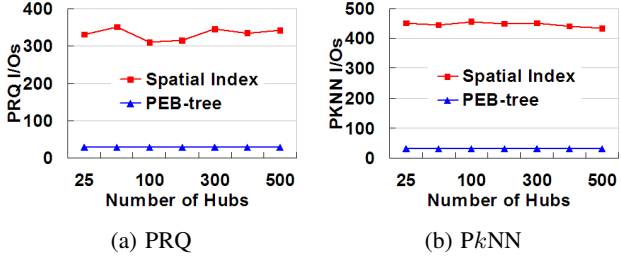

(a) PRQ  (b) P$k$NN

Figure 16: PEB-Tree vs. Spatial Index

## 7.8 Effect of Object Speed

We are also interested in studying how the object speed affects the query performance of both approaches. We vary the maximum speed from 1 to 6, choosing object speeds in the range from 0 to the maximum speed at random. As shown in Figure 17, the query cost of the spatial index increases slightly when objects move faster for both types of queries. This is because the query algorithm of spatial index needs to enlarge the query window according to the maximum object speed. The higher the speed, the larger the final search region becomes, yielding a higher cost. Compared to the spatial index, the PEB-tree has relatively stable performance. Although the PEB-tree shares the query window enlargement problem with the spatial index approach, the location constraints used in the PEB-tree are dominated by the policy compatibility, which significantly reduces the effect of this location-related parameter.

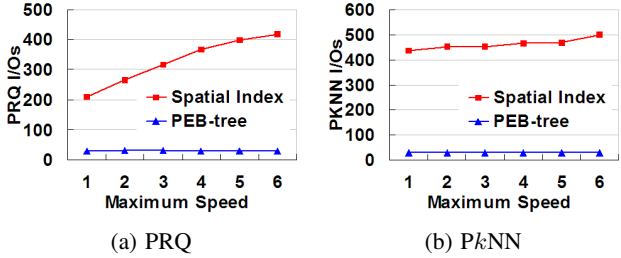

(a) PRQ  (b) P$k$NN

Figure 17: Effect of Object Speed

## 7.9 Effect of Updates

To observe the effect of updates on query performance, we measure the query costs if both approaches each time 25% of the data set has been updated. The experiments are conducted until the data set has been fully updated twice. The results, in Figure 18, show that the query cost of both approaches only fluctuates slightly. This is because the two indexes share the same base structure, i.e., the $B^x$-tree. The fluctuations are mainly caused by the amount of objects belonging to different time partitions in the trees.

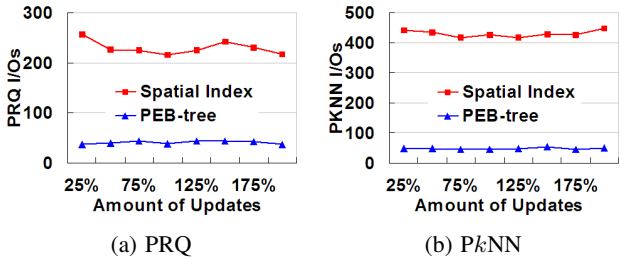

(a) PRQ  (b) P$k$NN

Figure 18: Effect of Updates

## 7.10 Cost Function Evaluation

We end by evaluating the accuracy of the cost function $C$ developed in Section 6. We compare the I/O cost as obtained from the cost function $C$ with the actual I/O cost. The comparison is conducted by varying one of three parameters at a time: the total number of users, the number of policies per user, and the grouping factor. We consider these three parameters because they are the main factors that affect the query performance of the PEB-tree, as shown in the previous experiments. The results are shown in Figure 19. From all the figures, we can see that the estimated cost tracks the actual cost quite well.

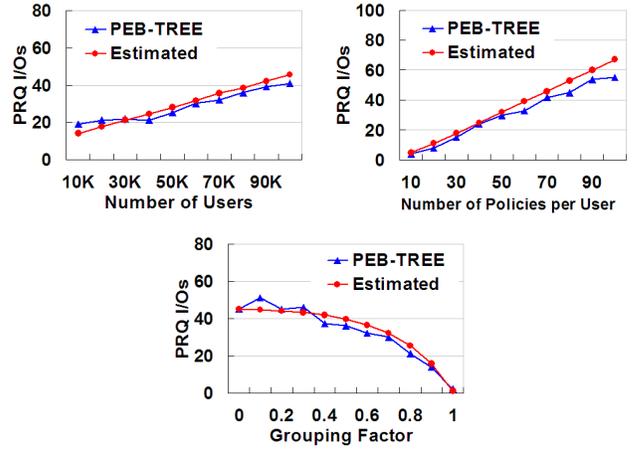

Figure 19: Cost Function Evaluation

## 8. CONCLUSIONS AND FUTURE WORK

We consider the problem of efficiently supporting range and $k$ nearest neighbor queries in a setting that affords moving users of location-based services peer-wise location privacy. Specifically, different peer users are allowed to see the location of a user when the user is within a specified spatio-temporal range.

To support the resulting privacy-aware queries, we present a new indexing technique, called the PEB-tree, that leverages the $B^x$-tree that is based on the $B^+$-tree. This is enabled by a technique that encodes both the location privacy compatibility and the spatial proximity among users in a one-dimensional value that is amenable to $B^+$-tree indexing; thus, users who tend to be allowed to see each others' locations and who are spatially close tend to be stored together on disk. Range and $k$ nearest neighbor query algorithms are presented that exploit the PEB-tree to simultaneously filter candidate users according to both privacy compatibility and spatial proximity.

An empirical performance study compares the proposed techniques with an existing approach that uses simply a spatial index, and the study offers insight into the behavior of the proposed techniques for wide variety of workloads. The study shows that the proposals outperform the existing approach very substantially.

Several directions for future research exist. It is relevant to consider multiple policies between two users for computing policy compatibility degree. Similarly, it is relevant to explore new encoding and accompanying querying techniques. Moreover, it is of interest to extend other types of location-based queries to take into account peer-wise privacy concerns.

## Acknowledgments

C. S. Jensen was supported in part by Geocrowd, an Initial Training Network under FP7 - People Marie Curie Actions, funded by the European Commission.